\documentclass[12pt]{iopart}
\usepackage{iopams}
\usepackage{harvard}
\usepackage{graphicx}
\begin{document}


\title[FEM, DG, and FD Evolution Schemes in Spacetime]{Finite Element, Discontinuous Galerkin, and Finite Difference Evolution Schemes in Spacetime}

\author{G Zumbusch}

\address{Institut f\"ur Angewandte Mathematik, 
Friedrich-Schiller-Universit\"at Jena,
07743~Jena, Germany}
\ead{gerhard.zumbusch@uni-jena.de}
\begin{abstract}
  Numerical schemes for Einstein's vacuum equation are
  developed. Einstein's equation in harmonic gauge is second order
  symmetric hyperbolic. It is discretized in four-dimensional
  spacetime by Finite Differences, Finite Elements, and Interior
  Penalty Discontinuous Galerkin methods, the latter related to Regge
  calculus. The schemes are split into space and time and new
  time-stepping schemes for wave equations are derived. The methods
  are evaluated for linear and non-linear test problems of the
  Apples-with-Apples collection.
\end{abstract}



\section{Introduction}

Numerical methods for the solution of Einstein's equation in general
relativity are mainly based on Finite Differences (FD) and
Pseudo-Spectral-Collocation \cite{meudon04,boyle} schemes space so
far. The Finite Element method (FEM), or more generally Galerkin
schemes have been used for reduced or auxiliary problems in numerical
relativity
\cite{Mukherjee98,Metzger04,Sopuerta06,Holst08,Hesthaven09}. However,
Galerkin methods are heavily used for the solution of wave problems in
areas like acoustic and electro-magnetic scattering and elastic waves
\cite{Cohen}. This is mainly due to their way to deal with
heterogeneous media and arbitrarily shaped geometric objects,
represented by unstructured grids. Furthermore, the convergence theory
of Galerkin methods is based on lower regularity (differentiability)
requirements than Finite Differences and spectral methods.

General relativity is governed by Einstein's equation, which can be
written as a system of second order partial differential equations in
spacetime. In order to define a well-posed initial-value (Cauchy)
problem, additional gauge conditions are needed. For the numerical
solution of the system, spacetime is usually split into space and time
$3+1$ and finally a time-stepping scheme is derived. Using a lapse-
and a shift-function, a sequence of space-like manifolds is
constructed, which fixes the gauge freedom. There are many
improvements of the original ADM \cite{ADM,York79} splitting like
BSSN \cite{bssn1,bssn2}. The equations are usually discretized in
space by FD or spectral schemes and independently in time by an
explicit integrator for ordinary differential equations.

The harmonic approach and its generalizations first incorporate the
harmonic gauge condition into Einstein's equation in spacetime to
derive a hyperbolic system
\cite{Fock59,Bruhat62,Reula98,Friedrich:2000qv,Pretorius05}. Afterwards,
the system is again split into space and time and
discretized. Generalized harmonic methods modify the gauge condition,
but usually preserve the hyperbolicity.

In this paper, we follow a slightly different approach. Starting with
the hyperbolic system of Einstein's equation in harmonic gauge, we
discretize first in spacetime. Introducing a global time-step, the
system is split afterwards in space and time. However, adaptive grid
refinement in space and local time-stepping schemes can also be
derived in a consistent way. This is similar to Regge calculus
\cite{Regge61,Sorkin75} in spacetime.

The main contribution of the paper however is the development of a
Finite Element and an Interior Penalty Discontinuous Galerkin (DG)
method for Einstein's vacuum equation. Both methods are derived from
a variational formulation, which is obtained from the Einstein-Hilbert
action and harmonic gauge. In fact, Galerkin methods are always based
on a variational version of the differential equations.

Galerkin schemes have been considered for the discretization of wave
equations in several ways so far: The wave equation
$\partial_{tt} u=\Delta u$ as an example problem is written
in variational form as
\[
\begin{array}{rcl}
\int_\Omega (\partial_{tt} u) w \, d^3x &=&
 -\int_\Omega (\nabla u)\cdot (\nabla w) d^3x ~\forall w 
\end{array}
\]
with trial functions $w$, integration over the spatial domain
$\Omega$, and zero boundary conditions.  This gives rise to FEM
\cite{dupont73,BakerBramble} and DG
\cite{Ainsworth06,grote06,Hesthaven09} in space schemes, used in
conjunction with a standard time integrator like the leapfrog
scheme. The first order in time formulation
$\partial_{t} v = \Delta u$ and $\partial_{t} u = v$ in variational
version in time reads as
\[
\begin{array}{rcl}
-\int_T  v (\partial_{t} w) dt &=& \int_T (\Delta u) w \, dt ~\forall w\\
-\int_T  u (\partial_{t} w) dt &=& \int_T v w \, dt ~\forall w
\end{array}
\]
on the interval $T$ and without initial value terms. In order to
obtain a time-stepping scheme, a time-discontinuous Galerkin method
can be constructed \cite{Jamet78,ErikssonJohnsonThomee85}. Note that
time continuous functions do not lead to a time-stepping scheme, but a
single large equation system for all times. We can combine both
Galerkin schemes to a spacetime FEM like
\[
\begin{array}{rcl}
\int_{\Omega \times T}  v (\partial_{t} w) dt \, d^3x &=&
 \int_{\Omega \times T} (\nabla u)\cdot (\nabla w) dt \, d^3x ~\forall w\\
-\int_{\Omega \times T}  u (\partial_{t} w) dt \, d^3x &=&
 \int_{\Omega \times T} v w \, dt \, d^3x ~\forall w    ~,
\end{array}
\]
continuous \cite{FrenchPeterson,AndersonKimn07} and discontinuous
\cite{HulbertHughes90,MonkRichter} in time.  In this paper, however,
we will consider second order in space and time formulations of type
\begin{equation}
\begin{array}{rcl}
\int_{\Omega \times T} (\partial_{t} u) (\partial_{t} w) dt \, d^3x &=&
 \int_{\Omega \times T} (\nabla u)\cdot (\nabla w) \, dt \, d^3x ~\forall w ~,
\end{array}
\label{eq:var0}
\end{equation}
again without boundary and initial value terms. It can be re-written
covariant and leads to time-stepping algorithms even for
time-continuous Galerkin discretizations, which differ from first
order formulations in general.

The first result of the paper in section~\ref{sec:equation} is in fact
the derivation of such a variational formulation of Einstein's
equation from the Einstein-Hilbert action. In addition, a linearized
formulation is discussed.

If we restrict the solution and trial functions in (\ref{eq:var0}) to
some finite dimensional spaces, we obtain Galerkin discretizations in
section~\ref{sec:scheme}.  Although the spacetime formulation relates
values at different points in space and time, it reduces to a
time-stepping scheme for global time steps.  The FEM scheme reduces
further to the leapfrog time-stepping for piecewise linear functions
in time, equidistant time steps, and without mixed
space-time-derivatives. Note that leapfrog is related to the
St\"ormer-Verlet scheme and a special case of the Newmark
scheme. However, in the general spacetime case the FEM and the
symmetric and non-symmetric DG spacetime schemes seem to be new. They
form the next result of this paper, see sections \ref{sec:fem} and
\ref{sec:ipdg}.

While the leapfrog scheme is explicit for FD in space, see
\cite{Cohen} and \cite[App. B]{Pretorius05}, the FEM method in space
requires the solution of a global equation system with mass matrix
$\int_\Omega u w \, d^3x$ each time step. The DG method in space is
computationally more efficient than FEM in general, because the mass
matrix is block-diagonal and the equation systems are easier to
solve. However, by a special choice of numerical quadrature rules
(mass-lumping) in FEM, see \cite{Cohen}, and a choice of orthogonal
ansatz functions in DG, see \cite{Riviere}, the mass matrix is
diagonal and the equation systems are trivial to solve.

Now we put together the variational formulation of Einstein's equation
and the spacetime Galerkin schemes and we obtain in
section~\ref{sec:schemeeinst}, as the main result, a FEM, a symmetric
and a non-symmetric Interior Penalty DG method for Einstein's full
vacuum equation. As an intermediate step we briefly discuss a
simpler, linearized version of Einstein's equation.

Memory requirements for nodal FD and piecewise linear FEM schemes for
Einstein's equation are comparable, namely ten metric component
values per grid node. The DG methods need this storage of $10$ values
for each element and each ansatz function, i.e. $10\cdot5$ or
$10\cdot16$ for linear or multi-linear functions, thus are more memory
intensive. The fields are needed for two previous and the current
time-slice in the leapfrog time-stepping.  We put the discrete fields
into the variational formulation, which now translates to non-linear
equation systems. The matrix entries are computed by numerical
quadrature rules. Additional storage may be required for the matrices
and solution of the equation systems, which depends on the solver.

Finally, some numerical experiments inspired by the Apples-with-Apples
test suite \cite{applesapples,applesapples2} are used to compare both
schemes with a more traditional FD scheme in
section~\ref{sec:appl}. The Galerkin schemes with piecewise linear
functions on equidistant, cartesian grids show comparable CFL
conditions, comparable second order accuracy, similar (sometimes
opposite sign) dispersion second order in grid spacing, and comparable
second order accurate harmonic gauge conditions. The errors on
unstructured grids additionally depend on the orientation of the
elements with respect to the wave characteristics and element angle
conditions.

In order to solve realistic test cases in general relativity,
techniques to handle apparent horizons are needed. Standard techniques
include the puncture approach
\cite{puncture,movingpunct1,movingpunct2}, excision
\cite{Pretorius05}, and singularity avoiding slicing conditions.
Slicing would lead to a generalized harmonic gauge. Excision is
compatible with harmonic gauge and the excised domain can be
approximated by unstructured grids, which seems to be most promising.
Furthermore, the Galerkin schemes have to be generalized to higher
order, which is straightforward in space, but is more difficult in
time for stability reasons.


\section{Einstein's Vacuum Equation}
\label{sec:equation}

\subsection{Strong Formulation}
We start with the standard derivation of Einstein's equation via the
Einstein-Hilbert action defined by
\[
S := \int_\mathcal{M} R \sqrt{-g} d^4x 
\]
in the case of vacuum, in the notation of \cite{Straumann04}. We
consider it as a function of the metric tensor $g_{\alpha \beta}$ and
its derivatives. The Ricci tensor $R_{\alpha \beta}$ and the Ricci
scalar $R=g^{\alpha \beta} R_{\alpha \beta}$ contain up to second
order partial derivatives of $g_{\alpha \beta}$. We are looking for an
extremum of $S$. The variation of $S$ is
\begin{equation}
\delta S=\int_\mathcal{M} (R_{\mu \nu} -\frac{1}{2} g_{\mu \nu}R) (\delta g^{\mu \nu}) \sqrt{-g} d^4x ~,
\label{eq:var2}
\end{equation}
as long as the variation $\delta g^{\mu \nu}$ vanishes at the boundary
of the domain $\mathcal{M}$. Otherwise we obtain an
additional boundary term
\begin{equation}
\frac{3}{2} \int_{\partial \mathcal{M}} 
g^{\alpha \beta} g^{\mu \nu} (\partial_\nu \delta g_{\mu \beta} -  \partial_\beta \delta g_{\mu \nu})
 n^{(g)}_\alpha d^3x 
\label{eq:bcterm}
\end{equation}
which can be used later for boundary conditions using derivatives of
$g_{\mu \nu}$.  We rename the variation
\[
v^{\mu \nu} := \delta g^{\mu \nu} ~.
\]
The variational formulation reads as: We seek a solution $g_{\alpha \beta}
\in V_a$ such that $\delta S = 0$ for all $v_{\alpha \beta} \in V_t$
with appropriate ansatz and trial spaces.
Dirichlet boundary conditions on (parts of) $\partial \mathcal{M}$ can
be built into $V_a$ and $V_t$: The solution takes the Dirichlet values
and the trial functions vanishes there. Boundary conditions involving
derivatives require an additional boundary term
like~(\ref{eq:bcterm}).
The variational formulation translates to the strong formulation as $R_{\mu
  \nu} -\frac{1}{2} g_{\mu \nu}R = 0$ or in vacuum
\[
R_{\mu \nu} = 0 
\]
with appropriate boundary conditions. However, in order to obtain a
well posed initial-boundary-value or Cauchy problem, we need an
additional gauge condition. We choose the standard harmonic gauge with
\begin{equation}
\Gamma^\alpha := g^{\rho \sigma}\Gamma^\alpha_{\rho \sigma} = 0 ~,
\label{eq:harm}
\end{equation}
which is a condition on first order derivatives of $g_{\alpha
  \beta}$.
This way, we can modify Einstein's equation as
\begin{equation}
R^{(h)}_{\mu \nu} := R_{\mu \nu} - \frac{1}{2} g_{\alpha \nu} \partial_\mu \Gamma^\alpha - \frac{1}{2} g_{\alpha\mu} \partial_\nu \Gamma^\alpha = 0 ~,
\label{eq:strong}
\end{equation}
with principal part
\begin{equation}
R^{(h)pp}_{\mu \nu} := - \frac{1}{2} g^{\alpha
  \beta} \partial_\alpha \partial_\beta g_{\mu \nu}~.
\label{eq:ppstrong}
\end{equation}
Now, we have a quasi-linear, second order, symmetric hyperbolic
differential equation, which we will later discretize by finite
differences.
Note that this remains true if we switch to a generalized harmonic
gauge. Equation (\ref{eq:harm}) changes to $\Gamma^\alpha =
H^\alpha(x,g)$ with a gauge driver $H$. This driver may depend on
coordinates and the metric, but must be independent of derivatives of
$g$ in order to preserve the principal part $R^{(h)pp}$.

\subsection{Variational Formulation}
Galerkin discretizations are based on a variational formulation. We
start with the standard variational formulation (\ref{eq:var2}). By
Stokes' theorem, we can remove the second order derivatives.  With
harmonic gauge (\ref{eq:harm}) we arrive at a variational version of
(\ref{eq:ppstrong})
\begin{equation}
a(g,v) := \frac{1}{2}\int_\mathcal{M} g^{\alpha \beta} \sqrt{-g} \,(\partial_\alpha g_{\mu \nu}) (\partial_\beta v^{\mu \nu}) d^4x ~,
\label{eq:var1}
\end{equation}
which is symmetric in the first order derivatives of $g_{\mu \nu}$ and
$v^{\mu \nu}$ in the special case of a fixed background $g^{\mu
  \nu}$. Again there is an additional boundary term, if the variation
$v$ does not vanish on the boundary $\partial \mathcal{M}$
\begin{equation}
-\frac{1}{2} \int_{\partial \mathcal{M}}
g^{\alpha \beta} \sqrt{-g} \,(\partial_\beta g_{\mu \nu}) v^{\mu \nu}
    n^{(g)}_\alpha d^3x ~.
\label{eq:bc2term}
  \end{equation}
The remaining terms can be assembled in
\begin{equation}
\begin{array}{rl}
q(g,v) := \frac{1}{2}\int_\mathcal{M} g^{\alpha\beta} g^{\rho \sigma} \sqrt{-g} \Bigl( 
		   & (\partial_\alpha g_{\rho \mu}) (\partial_\beta g_{\sigma \nu}) - (\partial_\alpha g_{\rho \mu}) (\partial_\sigma g_{\beta \nu}) \\
		    + & (\partial_\alpha g_{\rho \mu}) (\partial_\nu g_{\beta\sigma}) + (\partial_\mu g_{\alpha\rho}) (\partial_\beta g_{\sigma \nu}) \\
                    -\frac{1}{2}& (\partial_\mu g_{\alpha \rho}) (\partial_\nu g_{\beta \sigma}) ~ \Bigr) \,v^{\mu \nu} d^4x ~,
\end{array}
\label{eq:var1q}
\end{equation}
which is quadratic and symmetric in the first order derivatives of
$g_{\mu \nu}$, compare also \cite[App. B]{Fock59}. The variational
formulation now reads as
\begin{equation}
  \begin{array}{rcl}
\mathrm{seek}~ g\in V_a ~\mathrm{such~that}~a(g,v)+q(g,v) &=& 0~ \forall v \in V_t \\
\mathrm{and}~ \Gamma^\alpha &=& 0 ~.
\end{array}
\label{eq:varcont}
\end{equation}
Note that metric $g \in V_a$ in (\ref{eq:varcont}) does not need to have well
defined second derivatives as in (\ref{eq:strong}) and may be chosen
in an appropriate Sobolev space.
In the case of a non vanishing energy-momentum tensor $T^{\mu \nu}$
additional terms of type
\[
b(g,v):=\int_\mathcal{M} 
(g_{\alpha \nu}g_{\mu \beta} - \frac{1}{2}g_{\mu \nu} g_{\alpha \beta})
 \sqrt{-g} \, T^{\alpha \beta} \, v^{\mu \nu} d^4x
\]
appear on the right-hand side of (\ref{eq:varcont}).

Different types of initial and boundary conditions can be imposed on
$\partial \mathcal{M}$ by standard procedures to define a Cauchy
problem: Homogeneous Dirichlet values are directly incorporated into
all functions in $V_a$ and $V_t$. Inhomogeneous Dirichlet conditions
are built into the solution $g$, either direct in the discrete
numerical scheme, or via an additive splitting into a homogeneous
auxiliary solution and a non-homogeneous function for the boundary
conditions.  Neumann boundary conditions and other conditions based on
derivatives of the solution on parts of $\partial \mathcal{M}$ lead to
additional terms in $a$ of type (\ref{eq:bc2term}), where
$\partial_\beta g_{\mu \nu}$ is replaced by the given derivatives. The
functions in $V_a$ and $V_t$ do not vanish there.  ``Natural''
boundary conditions can be defined as vanishing term
(\ref{eq:bc2term}), that is $g^{\alpha \beta} \sqrt{-g}
(\partial_\beta g_{\mu \nu}) n^{(g)}_\alpha=0$.  The conditions can be
translated back into a strong formulation via (\ref{eq:bcterm}).

\subsection{Linearized Equations}
\label{sec:linear}

In a weak field approximation of Einstein's equation, we 
neglect the first order derivatives in (\ref{eq:strong}) and arrive at
$R^{(h)pp}_{\mu \nu} =0$ for some background metric $\hat{g}^{\mu
  \nu}$. In the variational version (\ref{eq:varcont}), we can neglect
$q(g,v)$ and solve for $a(g,v)=0$ instead, again for a fixed background
metric $\hat{g}^{\mu \nu}$.
\begin{equation}
a(g,v) := \frac{1}{2}\int_\mathcal{M} \hat{g}^{\alpha \beta} \sqrt{-\hat{g}} \,(\partial_\alpha g_{\mu \nu}) (\partial_\beta v^{\mu \nu}) d^4x
  \label{eq:varcurve}
\end{equation}
The linearized version of the harmonic gauge condition (\ref{eq:harm}) reads
\begin{equation}
\hat{g}^{\alpha \beta} \hat{g}^{\mu \nu}(\partial_\mu g_{\nu \beta} -\frac{1}{2} \partial_\beta g_{\mu \nu})=0 ~.
\label{eq:lingaugeb}
\end{equation}

Now, we simplify the equations even further and consider a weak field in flat space. The linearization is taken around Minkowski metric
$\hat{g} = \eta:=\mathrm{diag}(-1,1,1,1)$ and we obtain the strong
formulation
\begin{equation}
  -\frac{1}{2} \square g_{\mu \nu} = 0
  \label{eq:stronglin}
\end{equation}
with $\partial^\alpha=\eta^{\alpha \beta}\partial_\beta$ and
$\square=\partial^\alpha \partial_\alpha$.  This translates to the
variational version
\begin{equation}
\mathrm{seek}~ g\in V_a ~\mathrm{such~that}~ a(g,v):=\frac{1}{2}\int_\mathcal{M} \eta^{\alpha \beta} (\partial_\alpha g_{\mu \nu}) (\partial_\beta v^{\mu \nu}) d^4x = 0~ \forall v \in V_t ~.
  \label{eq:varlin}
\end{equation}
The harmonic gauge condition (\ref{eq:lingaugeb}) reduces to
\[
\partial^\mu g_{\mu \nu} -\frac{1}{2} \eta_{\mu \nu} \eta^{\alpha \beta} \partial^\mu g_{\alpha \beta} = 0~,
\]
which can be further simplified by the substitution $h_{\mu \nu}:=g_{\mu \nu}-\frac{1}{2} \eta^{\alpha \beta} g_{\alpha \beta}$ to
\begin{equation}
   \partial^\mu h_{\mu,\nu} = 0~.
  \label{eq:gaugelin}
\end{equation}
The differential equation still is (\ref{eq:stronglin}) $-\frac{1}{2}
\square h_{\mu \nu} =0$, now with a divergence free $h$. The gauge
conditions are linear and can be incorporated into the spaces $V_a$
and $V_t$.


\section{Numerical Schemes}
\label{sec:scheme}
\subsection{Finite Differences (FD)}
\label{sec:fd1}

For illustration purposes, the first numerical spacetime scheme will
be based on finite differences. We consider the discretization of
a linear, scalar, second order wave equation $-\square u = 0$ with
suitable initial and boundary conditions. On a one-dimensional,
equidistant grid with grid spacing $h$, we choose the stencil
$(u(x-h)-2u(x)+u(x+h))/h^2$, also abbreviated as $[1\, -2 ~ 1]/h^2$, to
approximate the second derivative. It is second order accurate for $u$
smooth enough.  The d'Alembert operator can be obtained by an
application of the stencil along each coordinate axis on a cartesian
grid.  The two dimensional stencil at a grid point $(i,j)$ for example
is
\[
\frac{u_{i-1,j}-2u_{i,j}+u_{i+1,j}}{h_0^2} - \frac{u_{i,j-1}-2u_{i,j}+u_{i,j+1}}{h_1^2} = 0 ~,
\]
which gives the explicit time stepping scheme
\[
u_{i+1,j} = 2u_{i,j}-u_{i-1,j}+ \left( \frac{h_0}{h_1} \right)^2 (u_{i,j-1}-2u_{i,j}+u_{i,j+1})
\]
using values at time slices $i-1$ and $i$ to calculate the values at
time slice $i+1$. This is the leapfrog scheme in time and can be
written as
\begin{equation}
u_{i+1} = 2 u_i-u_{i-1} + (h_0)^2 \Delta_h u_i
\label{eq:leapfrogfd}
\end{equation}
with a FD approximation of the spatial derivatives $\Delta$.  Note
that a CFL condition $h_0/h_k < 1$ for all $k>0$ must hold for
stability reasons \cite{Cohen}. The initial conditions can be
prescribed at two times slices $x_0 = 0$ and $x_0 = h_0$, the boundary
values at $x_k=0$ and $x_k=1$. Modifications for other types of
initial and boundary conditions do exist.

\subsection{Compact Finite Difference Stencils (FDM)}
\label{sec:fdm}

In order to generalize the FD stencils to mixed first and second order
derivatives, we consider an alternative construction. In the one
dimensional case, first derivatives can be approximated by central
stencils $u'(x+h/2) \approx (u(x+h)-u(x))/h$ at grid points
$x+h/2$. The second derivative can be calculated as a central stencil
of first derivatives $u''(x) \approx (u'(x+h/2)-u'(x-h/2))/h$ which
reduces to the one-dimensional FD stencil.  However, in two (and more)
dimensions the construction differs, if we consider cell-centered
first derivatives: We differentiate in one directions and average in
the other direction(s):
\[
\partial_0 u_{i+1/2,j+1/2} \approx \frac{1}{2} \left(\frac{u_{i-1,j}-u_{i,j}}{h_0} + \frac{u_{i-1,j+1}-u_{i,j+1}}{h_0} \right) ~.
\]
We obtain the second derivatives as stencils
\[
\partial_0 \partial_0 \approx \frac{1}{(2h_0)^2}
\left[
\begin{array}{ccc}
  -1 & 2 & -1 \\
  -2 & 4 & -2 \\
  -1 & 2 & -1 \\
\end{array}
\right]
~ \mathrm{and} ~
\partial_0 \partial_1 \approx \frac{1}{4h_0h_1}
\left[
\begin{array}{ccc}
  -1 & 0 & 1 \\
  0 & 0 & 0 \\
  1 & 0 & -1 \\
\end{array}
\right] ~.
\] 
The discretization of the d'Alembert operator again gives a
time-stepping scheme for time slice $i+1$. However, the scheme is no
more explicit like (\ref{eq:leapfrogfd}). Let us write the
difference stencil $[1~ 2~ 1]/4$ as the matrix $M$ and the
stencil $[-1~ 2\, -1]/(h_1)^2$ as matrix $A$. We obtain the scheme
\begin{equation}
M u_{i+1} = 2 M u_i - M u_{i-1} - (h_0)^2 A u_i ~.
\label{eq:leapfrog}
\end{equation}
We can compute the values $u_{i+1}$ at time slice $i+1$ by the
solution of a linear equation system with matrix $M$ using the values
$u_i$ and $u_{i-1}$ at time slices $i$ and $i-1$. The matrix is
positive definite, symmetric, and of bounded condition number. Hence,
the system is easy to solve numerically for large systems by standard
iterative solvers. Again, the CFL condition limits the time step size
$h_0$.

\subsection{Finite Element and Petrov-Galerkin Methods (FEM)}
\label{sec:fem}

We start with a variational version of the d'Alembert operator
(\ref{eq:var0}), a first step towards~(\ref{eq:varcurve}):
\begin{equation}
\frac{1}{2} \int_\mathcal{M} \eta^{\alpha \beta} \,(\partial_\alpha u) (\partial_\beta v) d^4x = 0 ~ \forall v
\label{eq:vardalembert2}
\end{equation}
Following standard procedures in FEM, we choose a set of global,
continuous, piecewise polynomial ansatz and trial functions $\tilde{\phi}_i
\in V_a$ and $\tilde{\psi}_j \in V_t$ as a basis of finite dimensional spaces
$V_a$ and $V_t$, and obtain a finite element method: Find the
coefficients $\tilde{u}_i$ of the solution $u = \sum_i \tilde{u}^i
\phi_i \in V_a$, such that (\ref{eq:vardalembert2}) holds for all
trial functions $v \in V_t$. This can be written in basis functions as
\begin{equation}
\frac{1}{2}\sum_i \tilde{u}^i \int_\mathcal{M} \eta^{\alpha \beta} \,(\partial_\alpha \tilde{\phi}_i) (\partial_\beta \tilde{\psi}_j) d^4x = 0 ~ \forall j 
\label{eq:varlinscal}
\end{equation}
and in matrix notation $\tilde{A}\tilde{u}=0$ with solution vector
$\tilde{u}$ and matrix $\tilde{A}=(\tilde{a}_{ij})$
\[
\tilde{a}_{ij}= \frac{1}{2}\int_\mathcal{M} \eta^{\alpha \beta} \,(\partial_\alpha \tilde{\phi}_i) (\partial_\beta \tilde{\psi}_j) d^4x ~.
\]
This is a spacetime discretization. Introducing a global time step,
we split functions
$\tilde{\phi}_i(x)=\phi^0_i(x_0)\phi^s_i(x_1,x_2,x_3)$ and
$\tilde{\psi}_j$, and the domain $\mathcal{M}=T\times \Omega$ into
time and space. Further, mixed space-time derivatives
$\eta^{a0}=\eta^{0b}=0$ do not occur with space index $a$, $b$. We
obtain
\[
\begin{array}{rcl}
\tilde{a}_{ij} &=& \frac{1}{2} 
\left( \int_T \eta^{0 0} (\partial_0 \phi^0_i) \,(\partial_0 \psi^0_j) dt \right)
\left(\int_\Omega \phi^s_i \psi^s_j d^3x \right) \\
&+& \frac{1}{2} 
\left(\int_T \phi^0_i \psi^0_j dt \right)
\left(\int_\Omega \eta^{a b} (\partial_a \phi^s_i) \, (\partial_b \psi^s_j) d^3x \right) ~.
\end{array}
\]
We introduce the mass matrix $M$ and matrix $A$ by
\[
  \begin{array}{rcl}
 m_{ij}&=&\frac{1}{2}\int_\Omega \phi^s_i \psi^s_j \, d^3x ~\textrm{and} \\
 a_{ij}&=&\frac{1}{2}\int_\Omega \eta^{a b} (\partial_a \phi^s_i) \, (\partial_b \psi^s_j) d^3x ~.
\end{array}
\]
In order to solve a Cauchy problem with initial conditions, we deviate
from standard FEM for self-adjoint problems in a single detail: In
order to mimic the behavior of the spacetime FD schemes, we start with
initial data at two time slices $i-1$ and $i$ and use the scheme to
calculate the next time slice $i+1$.  We use piecewise linear
functions $\phi^0_i(t)=\mathrm{max}(1-|t - x_{i}|/h_0 ,0)$ and
$\psi^0_j$ in time for equidistant time-steps $h_0$ and obtain the
system in time $\frac{1}{h_0}[-1 ~ 2 ~-1 ] M + \frac{h_0}{6} [1~4 ~1]
A$, which is of leapfrog type
\begin{equation}
\left( M-\frac{h_0^2}{6}A \right) u_{i+1} = \left(M +\frac{2h_0^2}{3}A \right)u_i - \left(M-\frac{h_0^2}{6}A \right) u_{i-1} ~.
\label{eq:leapfem}
\end{equation}
In the $1+1$ spacetime case, piecewise linear functions on an
equidistant space grid, we further obtain $M=\frac{h_1}{6} [1~4 ~1]$ and
$A=\frac{1}{h_1}[-1 ~ 2 ~-1 ]$.

The method can be interpreted as a Petrov-Galerkin method with
different ansatz $V_a$ and trial $V_t$ spaces: We use piecewise
polynomial functions centered at a grid point $i$ at time $i_0$ and
space location $(i_1,i_2,i_3)$ for a cartesian grid. The functions are
chosen piecewise linear in time. Let the grid points be in the time
domain $(0,k)$ with initial conditions at $i_0=0$ and $i_0=1$. We
compute the solution for all ansatz functions $\phi_i$ located at
times $(2,k)$. However, we choose the trial functions $\psi_j$
located at times $(1,k-1)$. The trial functions lag behind one time
slice, but are identical in space. This is exactly the idea to solve
for the next time slice $i_0+1$ and coincides with a leapfrog scheme
for equidistant time steps.

The solution of equation systems is the most expensive part of the
time stepping procedure. The advantage of FD schemes for leapfrog
(\ref{eq:leapfrogfd}) is that mass matrix $M$ is the identity and no
equation systems have to be solved. However, there is a common
technique in FEM called ``mass lumping'' to obtain diagonal matrices
$M$, too: Integration in terms of type $\int_\Omega \phi^s_i \phi^s_j
d^3x$ and $\int_T \phi^0_i \phi^0_j dt$ are approximated by numerical
quadrature rules on each finite element. For piecewise (multi-) linear
functions, quadrature rules prove to be sufficient, which are based on
the function values at the element vertices only. This is the
trapezoidal rule on an edge and its generalizations to rectangles,
cubes, triangles and tetrahedra. The ansatz functions fulfill
$\phi^s_i(x_j)=\delta_{ij}$ and the mixed products $(\phi^s_i
\phi^s_j)(x_k)=\delta_{ik}\delta_{jk}=0$ for $i \not= j$ vanish on all
element vertices $x_k$. Hence, off-diagonal entries $m_{ij}$ vanish
and $M$ is in fact a diagonal matrix. We arrive at the computational
efficiency of an FD scheme (\ref{eq:leapfrogfd}), once the integration
is done.

Note that (\ref{eq:varlinscal}) defines spacetime FEM also for higher
order methods in space by piecewise polynomial functions $\phi^s_i$ and
$\psi^s_j$, by pseudo-spectral Galerkin schemes in space, on
unstructured grids in spacetime, and for adaptive grid refinement in
spacetime. The approach does not easily extend to higher order methods
in time due to a lack of stability of the respective time-stepping
schemes.

\subsection{Interior Penalty Discontinuous Galerkin Methods (DG)}
\label{sec:ipdg}

Again, we start with the variational problem
(\ref{eq:vardalembert2}). However, we choose piecewise polynomial
ansatz and trial functions $\phi_i$ and $\psi_j$, which are no longer
continuous over element boundaries. This leads to additional
terms. Consider a common face $e_{ij}:=\partial E_i \cap \partial E_j$
of two neighbor elements $E_i$ and $E_j$ and normal unit vector
$n^{ij}$ oriented from $E_i$ to $E_j$. We denote the average by $\{ u
\} := ((u|_{E_i}) + (u|_{E_j}))/2$ and the jump by $[u]:=(u|_{E_i}) -
(u|_{E_j})$ on the face $e_{ij}$. Let the volume of the face be
$|e_{ij}|$.  We split the integration over $\mathcal{M}$ of
(\ref{eq:varlin}) into the integration over elements $E_i$ and all
faces $e_{ij}$ of the grid.
\begin{equation}
  \begin{array}{rl}
a(u,v) &:= \frac{1}{2}\sum_i \int_{E_i} \eta^{\alpha \beta} (\partial_\alpha u) (\partial_\beta v) d^4x \\
&\,\, -\frac{1}{2}\sum_{i<j} \int_{e_{ij}} \{ \eta^{\alpha \beta} n^{ij}_\alpha \partial_\beta u \} [v]  d^3x \\
&\,\, -\frac{1}{2}\sum_{i<j} \int_{e_{ij}} [u] \{ \eta^{\alpha \beta} n^{ij}_\alpha \partial_\beta   v \}  d^3x \\
&\,\, +\frac{1}{2}\sum_{i<j} \frac{c_p}{|e_{ij}|^{c_e}} \eta^{\alpha \beta} n^{ij}_\alpha n^{ij}_\beta \int_{e_{ij}} 
 [u][v] d^3x = 0
\end{array}
  \label{eq:varsipdg}
\end{equation}
The first jump term is obtained by Stokes' theorem, the second is added
for reasons of symmetry of $a$, and the last term with penalty
parameters $c_p$ and $c_e$ weakly imposes inter-element continuity.
We have modified the penalty term, originally strictly positive for
elliptic operators, by $\{ \eta^{\alpha \beta} n^{ij}_\alpha
n^{ij}_\beta \}$ due to the indefiniteness of the bi-linear form.

We choose polynomial ansatz and trial functions on each element and
combine them without continuity to global functions $\phi_i$ and
$\psi_j$. They define a basis of the finite dimensional spaces $V_a$
and $V_t$. Find coefficients $\tilde{u}^i$ such that
\[
\sum_i \tilde{u}^i a(\phi_i,\psi_j) = 0~~ \forall j ~.
\]
The scheme is called the symmetric interior penalty discontinuous
Galerkin scheme (SIPDG).  Note that an opposite sign of the second
jump term leads to the alternative non-symmetric NIPDG scheme, in our
case with penalty $c_p=0$. Boundary conditions require modifications
of the terms with outer boundary faces, see \cite{Riviere}.

If we use linear polynomials along each coordinate axis on an
equidistant grid as before, we can calculate the difference stencils
explicitly. In two dimensions $n=2$ for example, we use the local
nodal basis $(1-x_0)(1-x_1)$, $x_0(1-x_1)$, $(1-x_0) x_1$, $x_0 x_1$
and shift and scale it to each element. Again we solve for time slice
$i+1$ using slices $i-1$ and $i$. However, now there are four degrees
of freedom per element instead of one per node. With a penalty term
$c_e=1$ and different constants $c_p$ in both directions, we obtain
\[
A_{1,0} = A_{-1,0}^* = 
\frac{h_1}{12 h_0}
\left(
\begin{array}{rrrr}
  2 & 1 & 0 & 0 \\
  1 & 2 & 0 & 0 \\
  -4 & -2 & 2 & 1 \\
  -2 & -4 & 1 & 2 \\
\end{array}
\right)
+
\frac{c_{p1}} {6}
\left(
\begin{array}{rrrr}
  0 & 0 & 0 & 0 \\
  0 & 0 & 0 & 0 \\
  2 & 1 & 0 & 0 \\
  1 & 2 & 0 & 0 \\
\end{array}
\right)
\]
\[
A_{0,-1} = A_{0,1}^* =
\frac{h_0}{ 12 h_1}
\left(
\begin{array}{rrrr}
  2 & -4 & 1 & -2 \\
  0 & 2 & 0 & 1 \\
  1 & -2 & 2 & -4 \\
  0 & 1 & 0 & 2 \\
\end{array}
\right)
+
\frac{c_{p0}} {6}
\left(
\begin{array}{rrrr}
  0 & 2 & 0 & 1 \\
  0 & 0 & 0 & 0 \\
  0 & 1 & 0 & 2 \\
  0 & 0 & 0 & 0 \\
\end{array}
\right)
\]
\[
A_{0,0} =
-\frac{c_{p0} h_1} {6}
\left(
\begin{array}{rrrr}
  2 & 0 & 1 & 0 \\
  0 & 2 & 0 & 1 \\
  1 & 0 & 2 & 0 \\
  0 & 1 & 0 & 2 \\
\end{array}
\right)
+\frac{c_{p1} h_0} {6}
\left(
\begin{array}{rrrr}
  2 & 1 & 0 & 0 \\
  1 & 2 & 0 & 0 \\
  0 & 0 & 2 & 1 \\
  0 & 0 & 1 & 2 \\
\end{array}
\right)
\]
and a 5-block scheme for the degrees of freedom in element at
time $i+1$ and position $j$
\begin{equation}
A_{1,0} u_{i+1,j} = A_{0,-1} u_{i,j-1} + A_{0,0} u_{i,j} + A_{0,1} u_{i,j+1} - A_{-1,0} u_{i-1,j} ~.
\label{eq:leapdg}
\end{equation}
%
%
%
Note that for each element a linear equation system $A_{1,0}$ needs to
be solved. It is of the size of number of ansatz functions, which is
cheaper to solve than the single large equation system for the
FEM. However, the amount of work can be further reduced: It is
possible to choose the local ansatz functions orthogonal with respect
to the bi-linear form such that $A_{1,0}$ is in fact diagonal or even
the identity and no systems need to be solved any more. This way, we
obtain an explicit time-stepping scheme like (\ref{eq:leapfrogfd}).

For a second order differential equation in time, we need two initial
conditions, like $u(0,x_1)$ and $\partial_0 u(0,x_1)$. This can be
converted into data on two initial time slices $i=0$ and $I01$.
However, for the DG schemes, we need an initial spacetime
approximation in elements at times slices $0$ and $1$. For a linear
ansatz in time direction, initial data is needed at least at the
beginning and end of both time slices, namely three initial
values. These can be computed with a start-up calculation.

\subsection{Linearized Einstein's Equation}

In order to solve linearized Einstein's equation
(\ref{eq:stronglin}) resp. (\ref{eq:varlin}), we can generalize the
scalar schemes for $\square u$, apply these to each component $g_{\mu
  \nu}$, and set the background metric to Minkowski $\hat{g}=\eta$.
The linear gauge condition (\ref{eq:gaugelin}) needs to be
fulfilled. Divergence-free initial data guarantees this for all times
in the continuous case. However, numerical errors will lead to a
violation of the gauge condition.  DG methods easily allow for locally
divergence-free ansatz functions on each element. In contrast, it is
difficult to implement globally divergence-free symmetric tensor
fields in FEM analogous to divergence-free vector fields for Maxwell's
equation, see \cite{Nedelec80,Nedelec86}.

In the case of a prescribed curved background metric, we have to solve
the linear, variable coefficient problem $R^{(h)pp}_{\mu \nu} =0$.
The FD stencils are no longer applicable and we switch to the compact
FDM stencils. The FEM implementation is based on the variational formulation
\begin{equation}
\frac{1}{2}\sum_i u^i \int_\mathcal{M} \hat{g}^{\alpha \beta} \sqrt{-\hat{g}} \,(\partial_\alpha \phi_i) (\partial_\beta \psi_j) d^4x = 0 ~ \forall j ~.
\label{eq:varlinscalcurve}
\end{equation}
The DG method now reads as
\begin{equation}
  \begin{array}{rl}
a(u,v) &:= \frac{1}{2}\sum_i \int_{E_i} \hat{g}^{\alpha \beta} \sqrt{-\hat{g}} \,(\partial_\alpha u) (\partial_\beta v) d^4x \\
&\,\, -\frac{1}{2}\sum_{i<j} \int_{e_{ij}} \{ \hat{g}^{\alpha \beta} \sqrt{-\hat{g}} \,n^{ij}_\alpha \partial_\beta u \} [v]  d^3x \\
&\,\, -\frac{1}{2}\sum_{i<j} \int_{e_{ij}} [u] \{ \hat{g}^{\alpha \beta} \sqrt{-\hat{g}} \,n^{ij}_\alpha \partial_\beta   v \}  d^3x \\
&\,\, +\frac{1}{2}\sum_{i<j} \frac{c_p}{|e_{ij}|^{c_e}} \int_{e_{ij}} \{ \hat{g}^{\alpha \beta} n^{ij}_\alpha n^{ij}_\beta \sqrt{-\hat{g}} \}
 [u][v] d^3x = 0
\end{array}
  \label{eq:varsipdgcurve}
\end{equation}
where we have generalized the penalty term to $\{ \hat{g}^{\alpha
  \beta} n^{ij}_\alpha n^{ij}_\beta \sqrt{-\hat{g}} \}$.  The matrices
 $M$ and $A$ now depend on the background metric $\hat{g}^{\alpha
  \beta} \sqrt{-\hat{g}}$, which varies in spacetime. Procedures to
construct a diagonal $M$ like mass-lumping in FEM in
section~\ref{sec:fem} and orthogonal ansatz functions in DG in in
section~\ref{sec:ipdg} have to be performed on a per-element basis and
are thus more expensive, as are procedures to construct
divergence-free ansatz spaces. Once the matrix entries have been
computed, the linear equations system of type (\ref{eq:leapfem}) and
(\ref{eq:leapdg}) can be solved by standard solvers.

\subsection{Einstein's Vacuum Equation}
\label{sec:schemeeinst}

We generalize the compact FDM stencils to Einstein's vacuum equation
(\ref{eq:strong}): The variable metric $g_{\mu \nu}$ and its second
order derivatives $R_{\mu \nu}^{(h)}$ are chosen node centered (at grid points),
but the first order derivatives $\Gamma^\alpha_{\beta \gamma}$ are
chosen cell centered. The inverse metric $g^{\mu \nu}$ is used to
calculate $\Gamma^\alpha_{\beta \gamma}$ and $\Gamma^\alpha$ and is
also cell centered, defined as the inverse of the cell average of the
metric $g_{\mu \nu}$. The products of averaged $\Gamma$ enter the
Ricci tensor, as well as the node centered derivatives of
$\Gamma$. This way, we can use the standard formulas
$\Gamma^\alpha_{\beta \gamma}:=\frac{1}{2}g^{\alpha
  \sigma}(\partial_\beta g_{\gamma \sigma}+\partial_\gamma g_{\beta
  \sigma} - \partial_\sigma g_{\beta \gamma})$, $R_{\mu
  \nu}:=\partial_\alpha \Gamma^\alpha_{\mu \nu} - \partial_\mu
\Gamma^\alpha_{\nu\alpha} + \Gamma^\beta_{\alpha \beta}
\Gamma^\alpha_{\mu\nu} - \Gamma^\beta_{\mu\alpha}
\Gamma^\alpha_{\nu\beta}$, (\ref{eq:harm}), and (\ref{eq:strong}) to
set up non-linear, discrete Einstein's equation and derive the
time-stepping scheme. Note that no code generated by a symbolic
algebra program is needed.

The FEM and DG Galerkin schemes can also be generalized to
Einstein's equation. The form $a$ (\ref{eq:var1})
resp. (\ref{eq:varsipdg}) and the quadratic term $q$ (\ref{eq:var1q})
define the variational problem (\ref{eq:varcont}a). The integration is
done numerically. The integral $\int_\mathcal{M}$ is split into
integrals over an element $\sum_i \int_{E_i}$ (and a face $e_{ij}$ in
(\ref{eq:varsipdg})). The integrals over a single element $E_i$ and
face $e_{ij}$ are approximated by a numerical quadrature rule. The
integrands of $a$ and $q$ are evaluated at the quadrature
points.

The matrices $M$ and $A$ now depend on the current metric $g^{\alpha
  \beta} \sqrt{-g}$ and the equation systems of type
(\ref{eq:leapfem}) and (\ref{eq:leapdg}) are non-linear.  The
time-stepping schemes are implicit and require the solution of a
non-linear equation system for each time-slice. The DG method leads to
a set of easy to solve local equation systems for each element. The
FDM and the FEM have global coupling of the degrees of freedom of a
time slice. In both cases standard non-linear solvers can be
used. Note that the explicit FD method both gives an initial guess for
a locally fixed background metric $g^{\mu \nu}$ and can be used as a
preconditioner for the principle part in an iterative solver.

The harmonic gauge condition (\ref{eq:harm}) now is a non-linear
condition and cannot be incorporated into a linear ansatz space
$V_a$. Note that a change of variables leads to a formulation of
Einstein's equation with a new metric $\mathfrak{g}^{\mu
  \nu}:=\sqrt{-g} g^{\mu \nu}$ and a linear gauge condition
$\partial_\mu \mathfrak{g}^{\mu \nu} = 0$, which could be built into
$V_a$.

Note that Regge calculus also discretizes a variational principle in
spacetime for simplicial grids \cite{Sorkin75}. It can be considered a
DG spacetime scheme with piece-wise constant metric tensor $g_{\mu
  \nu}$. This way, (\ref{eq:varsipdg}) generalizes it to higher order
and arbitrary element shapes. However, Regge calculus does not use
coordinates and is based on purely geometric entities like edge
lengths and defect angles. Furthermore, the variation is with respect
to the degrees of freedom, which are the squared edge lengths in Regge
calculus and values of the metric in (\ref{eq:varsipdg}).


\section{Applications}
\label{sec:appl}
\subsection{Linear  Plane Wave}

For illustration purposes, we perform some numerical experiments with
the schemes of section~\ref{sec:scheme}. The test cases are adapted
from the Apples-with-Apples test suite
\cite{applesapples,applesapples2}. We document and compare convergence
and stability of the schemes in different settings.

We start with a mono-chromatic traveling plane wave for linearized
Einstein's equation (\ref{eq:stronglin}) and (\ref{eq:varlin}) with
harmonic gauge (\ref{eq:gaugelin}).  We use periodic boundary
conditions and a Courant factor $1/2$. The one-dimensional (1+1) test
case is defined on the spatial unit interval $(0,1($. The exact
solution and initial data is $g_{00}=g_{11}=-g_{01}=\mathrm{sin}
2\pi(x_1-x_0)$. We use an equidistant grid and run all schemes of
sections~\ref{sec:fd1} to~\ref{sec:ipdg}. Note that the original
Apples-with-Apples tests were constructed for non-linear numerical
codes, such that very small wave amplitudes effectively ran the codes
in the regime of the linearized equations. Standard non-linear solvers
like Newton's method in this case reduce the problem to a linear
one. Hence, we directly ran a linear code for the linear problem. This
is why we can use arbitrary amplitudes of the solution rather than
very small ones \cite{applesapples}.

\begin{figure}[bt]
  \centering
  \hspace*{-0.9cm}
  \includegraphics[width=.53 \textwidth]{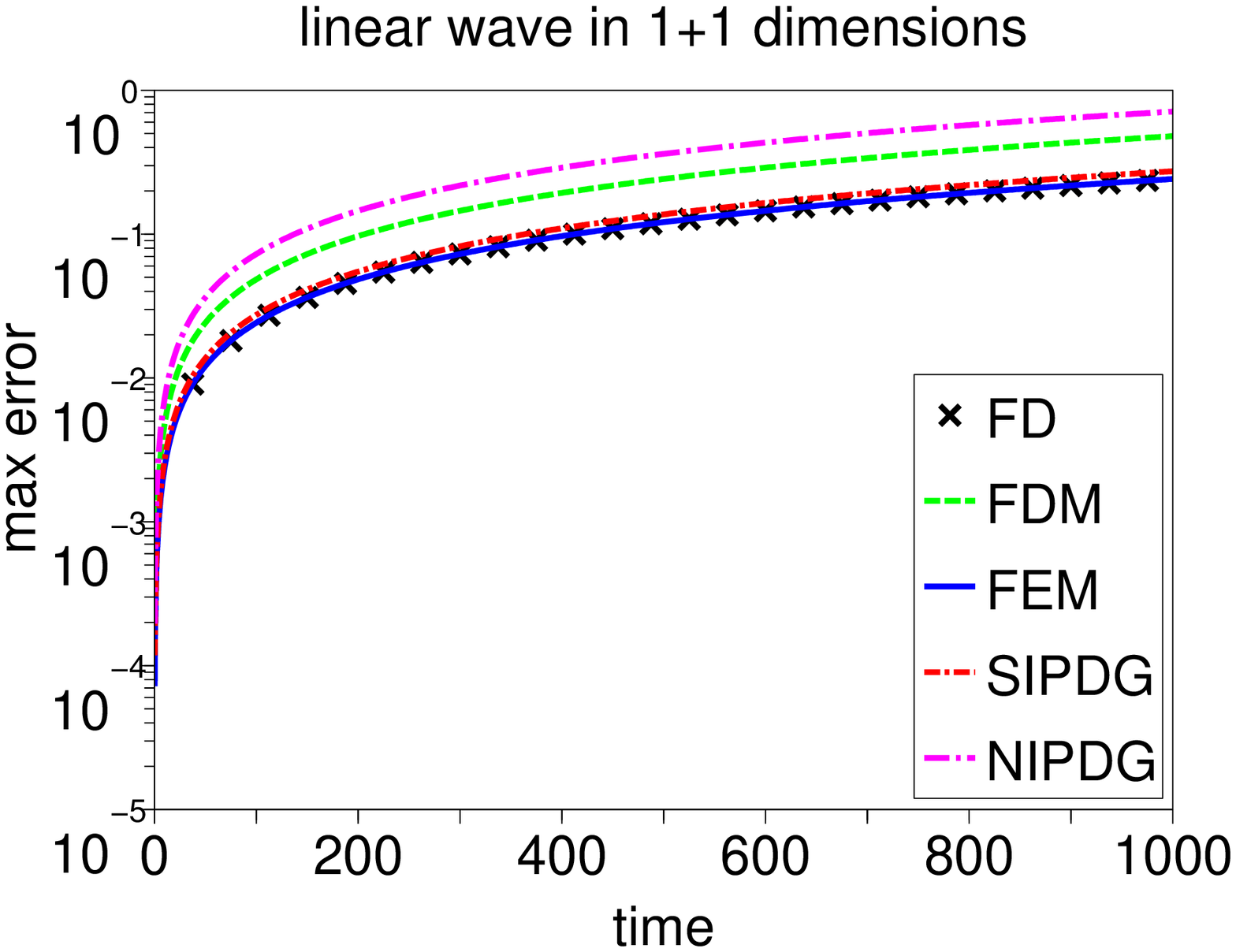}
  \hspace*{-0.7cm}
  \includegraphics[width=.53 \textwidth]{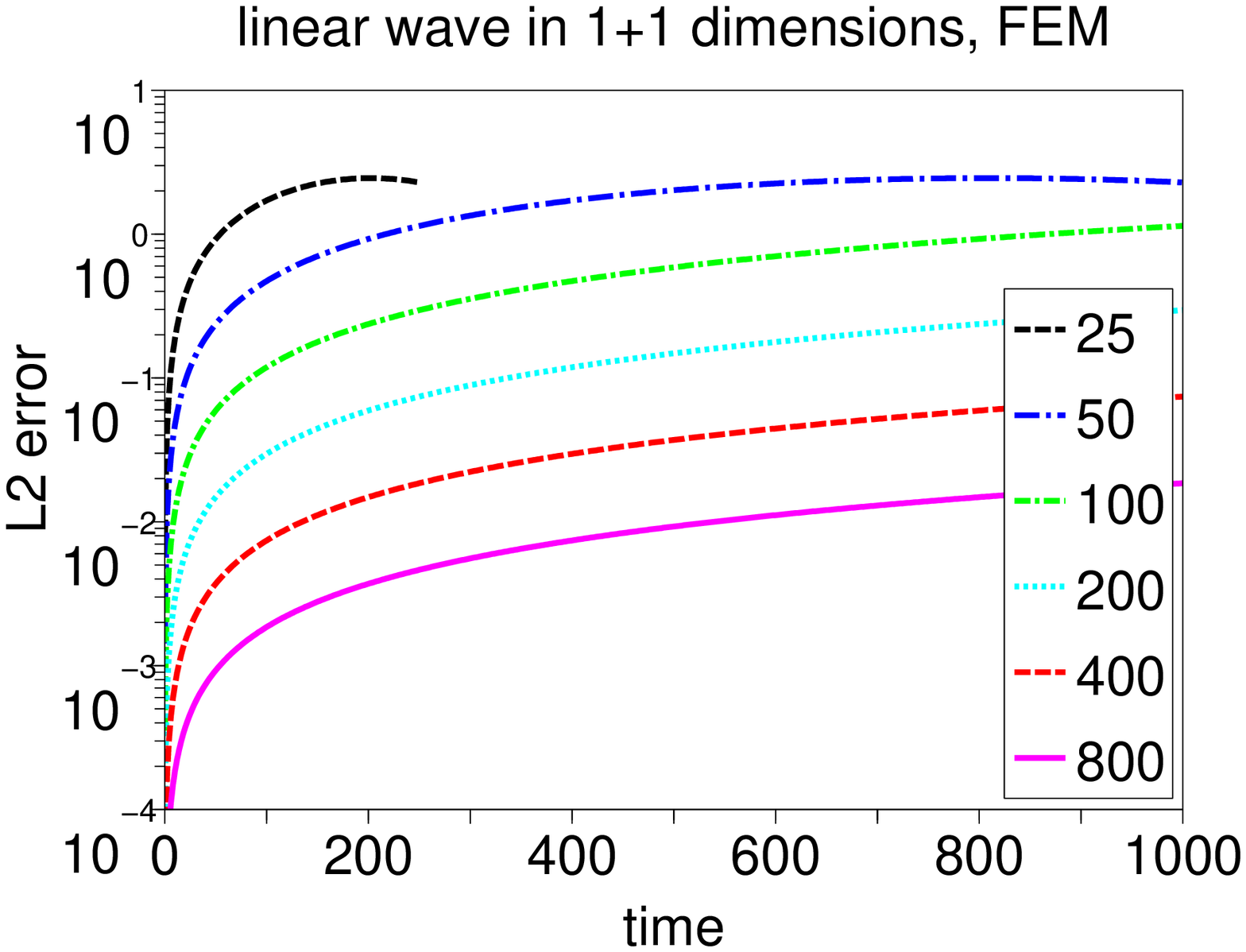}
  \hspace*{-1.6cm}
  \caption{1+1 linear plane wave: Time evolution of spatial maximum error with $h_1=1/200$ (left) and of the $l^2$-error of a FEM solution with $n=1/h$ (right).}
  \label{fig:lin1}
\end{figure}

\begin{figure}[bt]
  \centering
  \hspace*{-0.9cm}
  \includegraphics[width=.53 \textwidth]{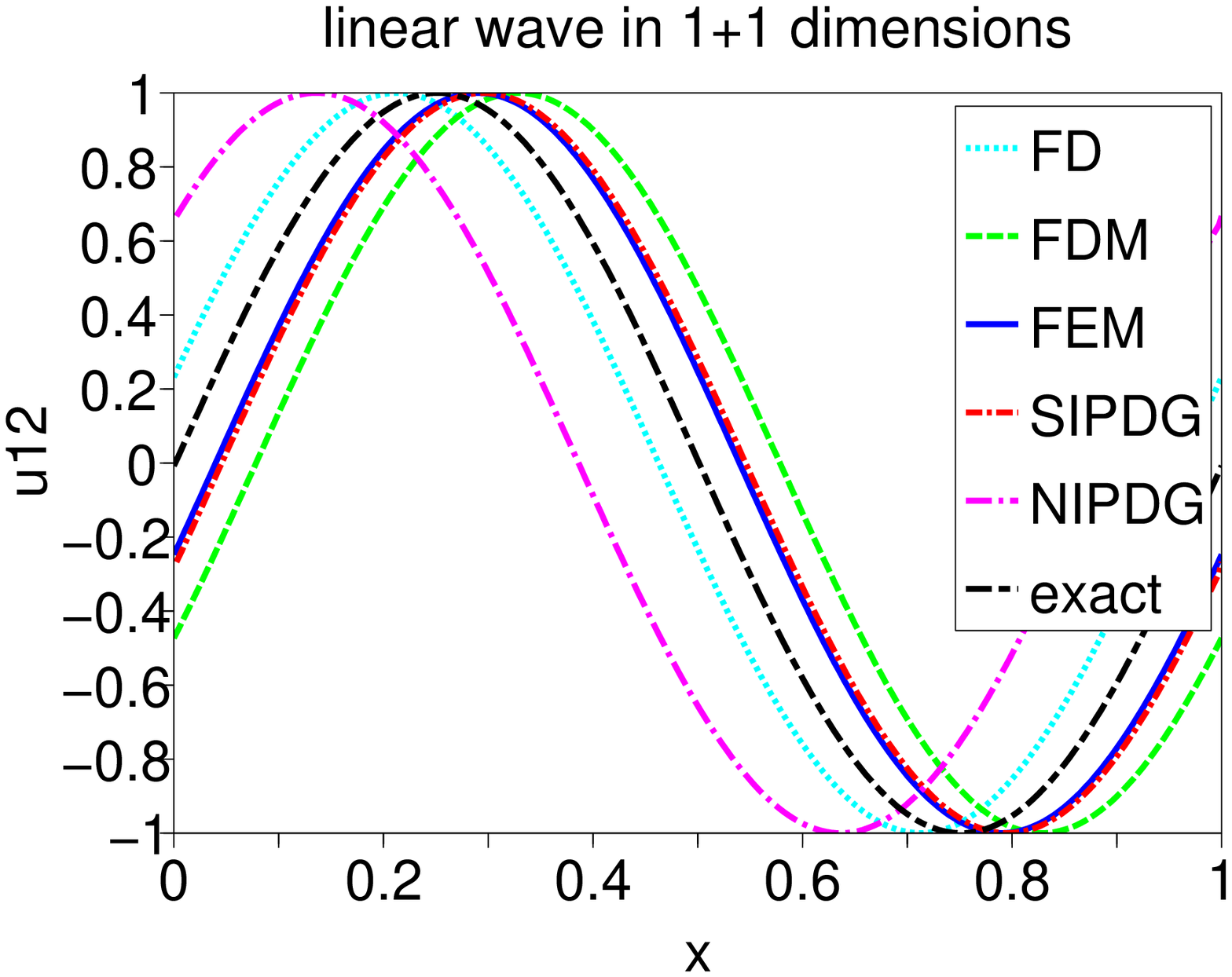}
  \hspace*{-0.7cm}
  \includegraphics[width=.53 \textwidth]{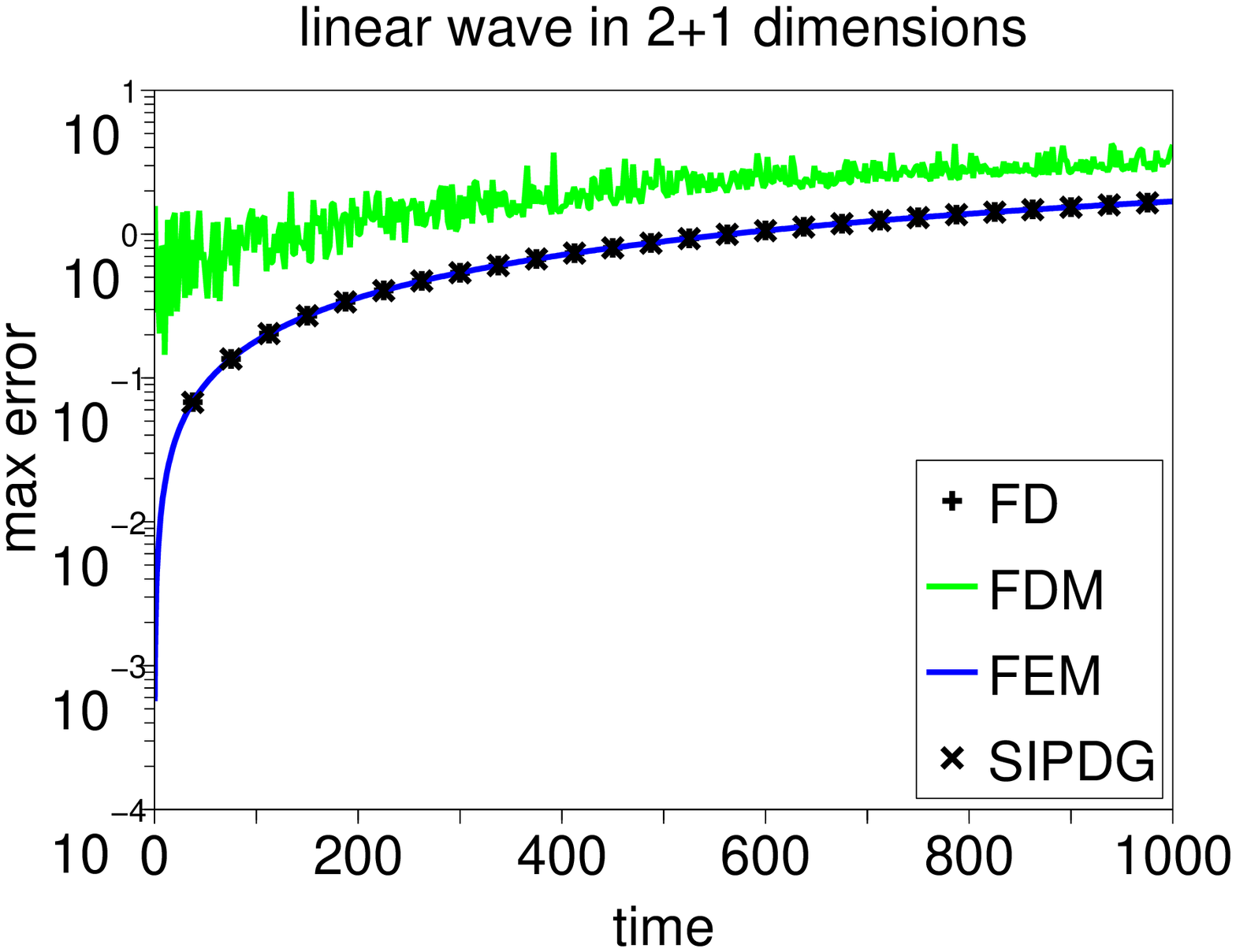}
  \hspace*{-1.6cm}
  \caption{Solution of a linear plane wave in 1+1 for $h_1=1/200$  at
    final time $x_0=1000$ (left). Evolution of the maximum error of a wave in 2+1 on a cartesian grid with $h_1=1/100$ (right).}
  \label{fig:lin1sol}
\end{figure}

In figure~\ref{fig:lin1} the time evolution of the spatial maximum
error at the grid points for a resolution of $h_1=1/200$ is depicted
for the FD, FDM, FEM, SIPDG and NIPDG. We use penalty parameters
$c_{p0}=1$ and $c_{p1}=2$ for SIPDG. Note that continuous error norms
like $L_2(0,1)$ more natural for FEM show a similar behavior with
exception of the very first time steps, where an additional
interpolation error is added to the global error. The point-wise
divergence is bounded, although we do not take any measures to control
it. This does not seem to be necessary. The solution in
figure~\ref{fig:lin1sol} (left) shows the spatial errors at the final
time $x_0=1000$. We see mainly dispersion and the phase error of the
different schemes, no errors in the amplitude. This is why the error
in fact even decreases after some time, see figure~\ref{fig:lin1}
(right) $n=25$ and $n=50$. We observe a second order convergence of the
error, the phase error and the divergence in $h_1$ for all schemes.

The two-dimensional (2+1) test case is defined on the spatial unit
square $(0,1(^2$ with periodic boundary conditions. The exact solution
and initial data is $g_{01}=g_{02}=g_{12}=\mathrm{sin}
2\pi(x_1+x_2-\sqrt{2}x_2)$, $g_{11}=g_{22}=(\sqrt{2}-1)g_{01}$, and
$g_{00}=\sqrt{2}g_{01}$. We run all schemes on cartesian equidistant
grids, see figure~\ref{fig:lin1sol} (right), except for the NIPDG
scheme for a lack of stability. The SIPDG penalty term is chosen as
$c_e=1/2$, more precisely $\frac{c_p}{|e_{ij}|^{c_e}}=1/h_0$.  The
second order convergence is comparable to the $1+1$ case.

\begin{figure}[tb]
  \centering
  \hspace*{-0.9cm}
  \includegraphics[width=.53 \textwidth]{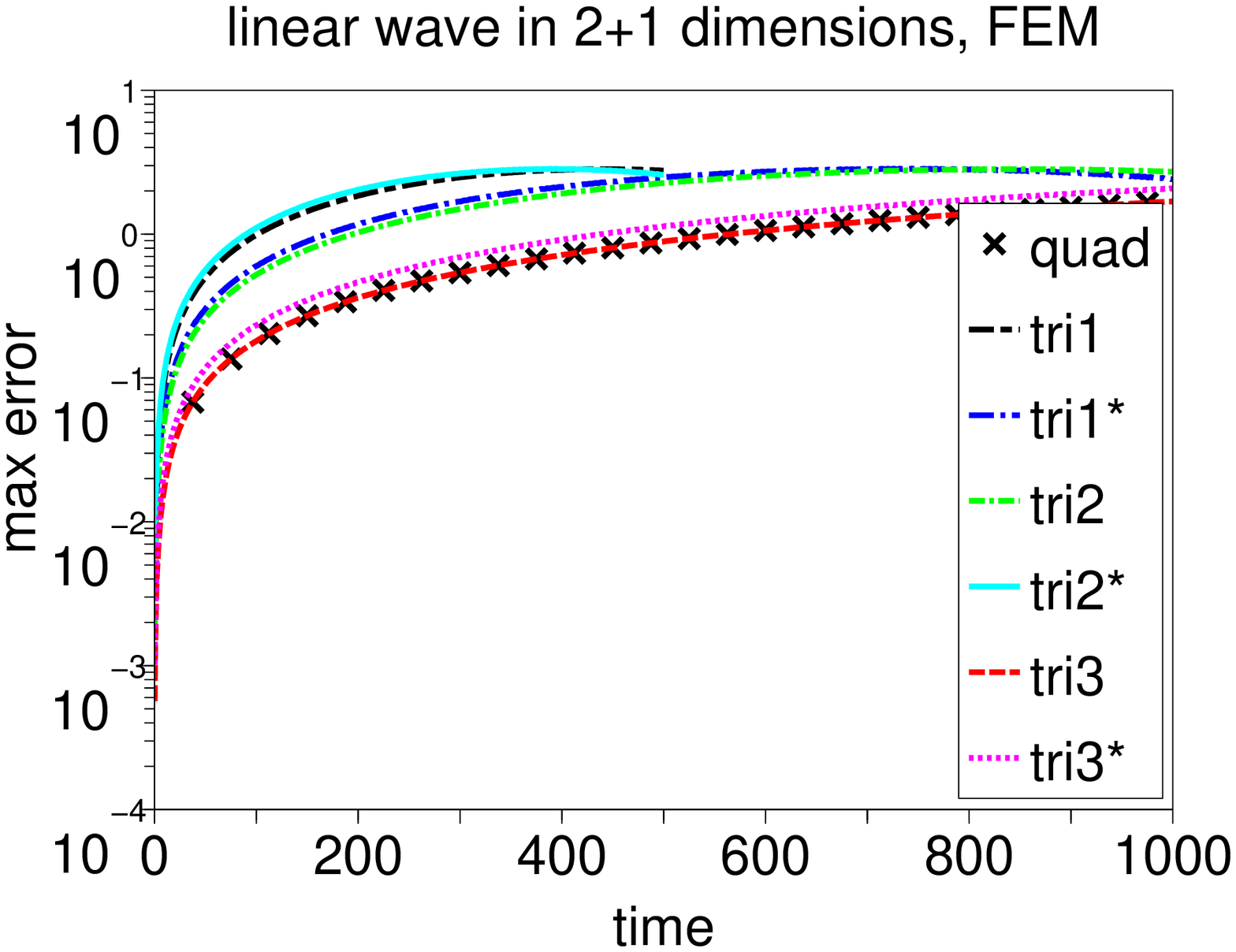}
  \hspace*{-0.7cm}
  \includegraphics[width=.53 \textwidth]{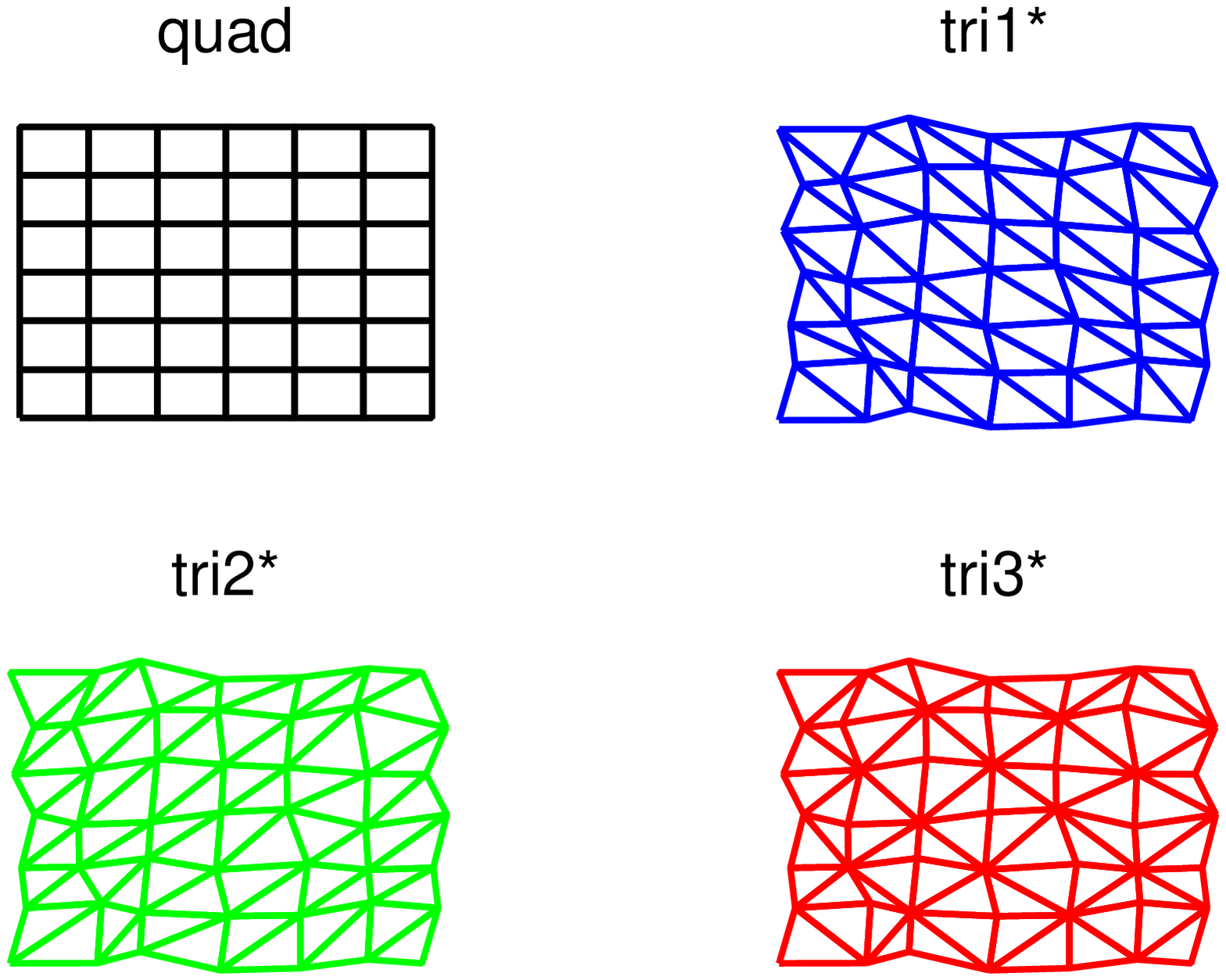}
  \hspace*{-1.6cm}
  \caption{Linear plane wave in 2+1 on some triangulations with $h_1=1/100$:
    evolution of the maximum error (left) and corresponding spatial grids (right).}
  \label{fig:lin2tri}
\end{figure}

In order to test the dependence on the spatial grid, we run the FEM
also on a number of triangular grids, both uniform (tri) and randomly
distorted (tri*), see figure~\ref{fig:lin2tri}. Now we obtain a strong
dependence of the error on the orientation of the elements. The
longest element edges tangential to the direction of the wave leads to
a larger approximation error than in normal direction or for quadratic
elements.

\subsection{Robust Stability Test for Linear Waves}

\begin{figure}[tb]
  \centering
  \hspace*{-0.9cm}
  \includegraphics[width=.53 \textwidth]{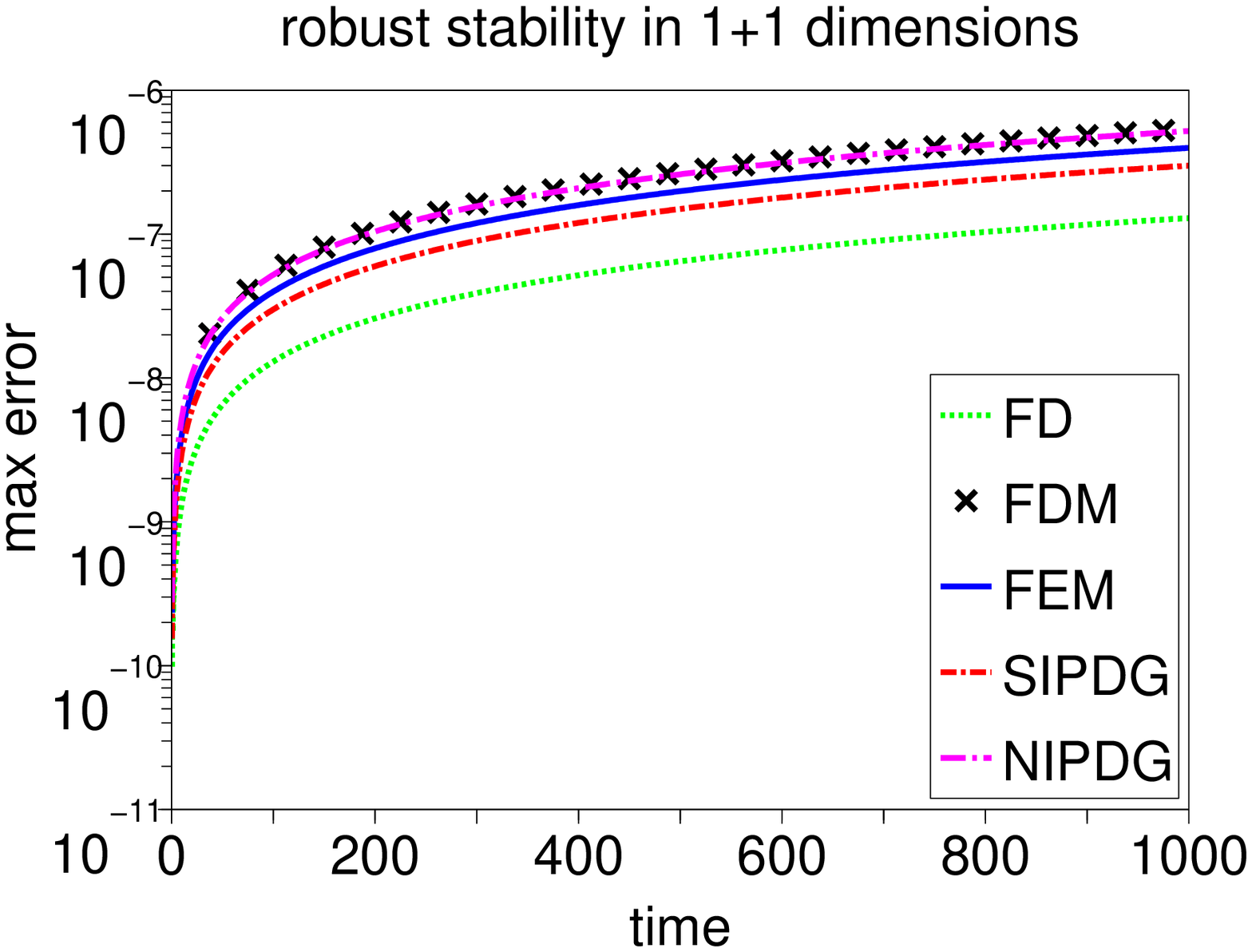}
  \hspace*{-0.7cm}
  \includegraphics[width=.53 \textwidth]{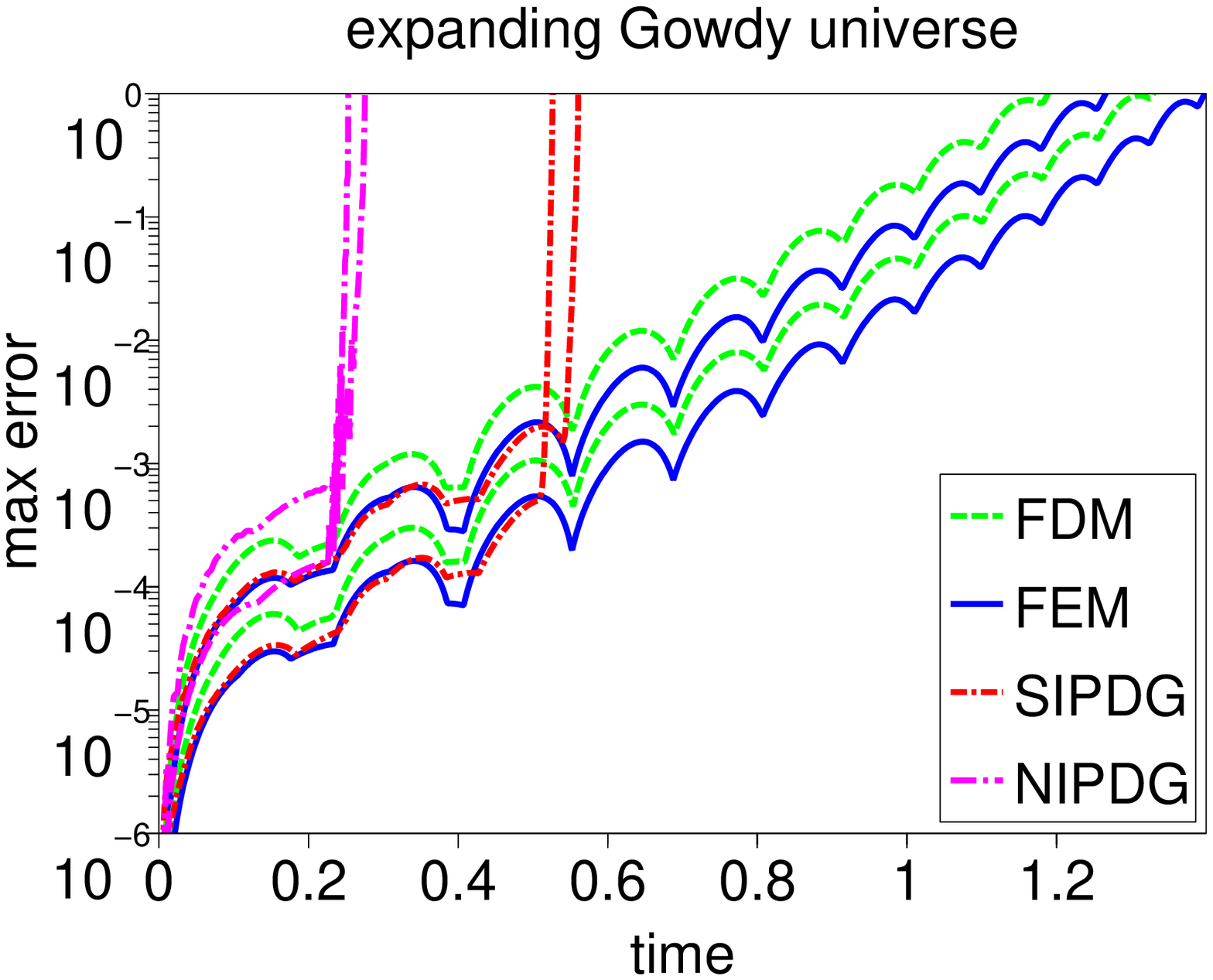}
  \hspace*{-1.4cm}
  \caption{Evolution of the maximum error of the linear robust stability test in 1+1 for $h_1=1/200$ (left) and of a non-linear Gowdy wave in 1+1 for $h_1=1/100$ and $h_1=1/200$ (right).}
  \label{fig:rob1}
\end{figure}

Now we consider a stability test for the linear wave equation. The
starting point is a random perturbation of the zero solution. We use
periodic boundary conditions on $(0,1($, equal distributed
$[-\epsilon,\epsilon]$ random values for all initial data, with
$\epsilon:=2.5\cdot 10^{-7} (h_3)^2$ according to
\cite{applesapples}.  In figure~\ref{fig:rob1} (left) we observe stability of
all schemes with oscillatory solutions for NIPDG and compact stencil
FDM.

\subsection{Nonlinear Polarized Waves in the Expanding Gowdy Universe}

The polarized Gowdy spacetime on the Torus $T^3$ is a model for a
gravitational wave in an expanding universe \cite{Gowdy71,New98}. We
use periodic boundary conditions on the spatial unit interval $(0,1($ in $x_3$
direction. The solution is constant along $x_1$ and $x_2$
direction. Since we use harmonic gauge, time axis $x_0$ differs from
\cite{applesapples}. We use a Courant factor $1/4$. The solution
$g_{\mu \nu}$ is given by
\[
\begin{array}{rcl}
g &=& \mathrm{diag}(-e^{(\lambda+3x_0)/2},~e^{x_0+p},~e^{x_0-p},~e^{(\lambda-x_0)/2}) ~\mathrm{with} \\
p &:=& \mathrm{J}_0(2\pi e^{x_0}) \mathrm{cos} (2\pi x_3 ) ~\mathrm{and} \\
 \lambda &:=& -2\pi e^{x_0} \mathrm{J}_0(2\pi e^{x_0}) \mathrm{J}_1(2\pi e^{x_0})  \mathrm{cos}^2 (2\pi x_3) 
- 2\pi \mathrm{J}_0(2\pi) \mathrm{J}_1(2\pi) \\
&& + 2 (\pi e^{x_0})^2(\mathrm{J}^2_0(2\pi e^{x_0}) + \mathrm{J}^2_1(2\pi e^{x_0}))
-\frac{1}{2} (2\pi)^2(\mathrm{J}^2_0(2\pi) + \mathrm{J}^2_1(2\pi))
\end{array}
\]

We run schemes of section~\ref{sec:schemeeinst} with 3rd order Gauss
quadrature (two points in each coordinate direction) on an
element. The SIPDG penalty terms are chosen as $c_{p0}=.5$ and
$c_{p1}=2$. In figure~\ref{fig:rob1} (right) we see the error for spatial
resolutions $h_1=1/100$ and $h_1=1/200$, which demonstrates second
order convergence. The DG methods do not seem to be as stable as the
others. However, many numerical schemes start to diverge at some time
$t$ due to the exponential growth of some of the solution components
\cite{applesapples}.


\section*{Conclusion}

We have developed new spacetime Finite Element (FEM) and Interior
Penalty Discontinuous Galerkin (SIPDG and NIPDG) schemes for second
order symmetric hyperbolic wave equations. The Discontinuous Galerkin
schemes are computationally more efficient, but require more memory
than FEM and Finite Differences methods. A variational formulation of
Einstein's equation in harmonic gauge was derived, based on up to
first derivatives of solution and trial functions. This led to new
Galerkin schemes for numerical relativity. The schemes were presented
and tested for second order accurate Galerkin schemes with
multi-linear functions and global time steps.  The Gowdy wave test
demonstrated the need for additional numerical stabilization. This
might be obtained by spatial filtering, artificial viscosity, or
streamline diffusion.

Extensions to arbitrary spacetime grids or (adaptive) local grid
refinement in spacetime are straightforward, but may lead to larger
and more expensive to solve equation systems. Higher order polynomials
or other more accurate (spectral) function spaces improve the spatial
accuracy of the schemes. However higher order in time schemes are more
difficult to construct.

\section*{Acknowledgments}

The author wants to thank G. Sch\"afer for several hints to the
literature. Furthermore, helpful comments by S. Husa and the anonymous
referees are acknowledged. This work was partially supported by DFG
grant SFB/TR7 ``gravitational wave astronomy''.


\section*{References}

\bibliographystyle{jphysicsB}
\bibliography{lin}

\end{document}